\documentstyle[12pt]{article}
\begin{document}{}
\begin{titlepage}
\begin{flushright}
gr-qc/9406045\\
QMW-Maths-1994-SPB-1 preprint \today\\[0.25in]
\end{flushright}
\begin{center}
{\Large\bf Hypertime Formalism for
Spherically Symmetric Black Holes and Wormholes}
{\\[0.5in] \large  Stephen P\@. Braham}
{\\[0.5in] \em       School of Mathematical Sciences \\
          Queen Mary and Westfield College, University of London \\
          Mile End Road. London E1 4NS \\
          United Kingdom \\
E-mail: S.P.Braham@qmw.ac.uk\\[0.5in]}
\end{center}

\centerline{\large\bf Abstract}
\vskip 0.25in
{\small

Recent work on an approach to the geometrodynamics of
cylindrical gravity waves in the presence of interacting scalar matter fields,
based on the Kucha\v{r} hypertime formalism, is extended to
the analogous spherically symmetric system. This produces a geometrodynamical
formalism for spherical black holes and wormholes
in which the metric variables are divided
into two classes, dynamical and redundant. The redundant variables
measure the embedding of a spacelike hypersurface into the spacetime,
and proper time in the asymptotically flat regions.
All the constraints can be explicitly solved for the momenta
conjugate to the embedding variables. The dynamical variables,
including an extra ADM mass for wormhole topologies, can
then be considered as functionals of the redundant ones, including
the proper time variable.
The solution of the resulting constraint system determines the momentum
conjugate to the proper time as a function of the other variables,
producing Unruh's Hamiltonian formalism for the spherical
black hole, whilst extending it to an arbitrary foliation
choice.
The resulting formalism is appropriate as a starting point for the
construction of a hypertime functional Schr\"odinger equation for
quantized spherically symmetric black holes and wormholes.

}
\end{titlepage}

\section{Introduction}\label{intro}
Classical and quantum geometrodynamics (CGD and QGD) are haunted, and made
more interesting, by the {\it problem of time}. This is simply
the problem that general relativity, through its general covariance,
does not possess a special, external, time parameter, and thus it is
difficult to understand the theory in terms of dynamics. In the classical
case, it is possible to avoid this problem if we view gravity through the
canonical formalism first set up by
Dirac \cite{DIRFIX}, as well as by Arnowitt, Deser and Misner (ADM)
\cite{ADM}. In this formalism, the spacetime 4-metric
is decomposed in terms of the familiar 3-metric $g_{ij}$ and the
corresponding lapse $N$, and shift $N^i$ variables, all introduced on
a specific foliation (prescribed by $N$ and $N^i$), labeled by a time
parameter $t$, and a spatial hypersurface position ${\bf x}$. The
action for the vacuum part of the
theory can then be written in the manifestly canonical form
\begin{equation}{}\label{hamac}
S = \int dt \, \int d^3x \left(
             {\pi}^{ij}({\bf x}) {\dot g}_{ij}({\bf x})
             - N ({\bf x}) {\cal H} ({\bf x})
             - N_i({\bf x}) {\cal H}^ ({\bf x}) \right),
\end{equation}
with the overdot representing partial differentiation with respect to $t$,
and $\pi^{ij}$ being the momenta conjugate to $g_{ij}$. This allows us
to view the theory as describing the evolving 3-geometry (and
corresponding momenta) on a hypersurface,
but under the superhamiltonian and supermomentum constraints
\begin{equation}{}\label{coneqs}
{\cal H} ({\bf x}) = 0,\ {\cal H}^i ({\bf x})=0,
\end{equation}
respectively, that are generated by variation of
$N$ and $N_i$ in Equation~(\ref{hamac}). These constraints contain
all the information to evolve the system, given the lapse and
shift \cite{ADM}.
In this sense, we can impose a concept of time on the theory. However,
even in the classical case, this position is a little uncomfortable,
as we are forced to consider each possible foliation separately, and,
furthermore, there is generally no geometric way of identifying a choice
of $t$. In the quantum theory, things are even worse. In this case,
the quantum equations describing the system are given by
Equation~(\ref{coneqs}), under the usual canonical quantization
rules for replacing phase-space variables with operators,
\begin{equation}{}\label{canquant}
g_{ij} \rightarrow \hat{g}_{ij} = g_{ij} \times,\ \pi^{ij}
\rightarrow \hat{\pi}^{ij} =
-i \frac{\delta}{\delta g_{ij}},
\end{equation}
operating on some wavefunction $\psi [g_{ij}]$.
These equations, as they stand, cannot
be viewed, in any way, as describing the evolution of a quantum state
with respect to $t$ and, in fact, describe a state that is manifestly {\it
independent of $t$}. So, our ``fix'' does not extend to the quantum
theory.

One possible solution to the problem of time is to view the constraints
given by Equation~(\ref{coneqs}) as representing evolution with
respect to other variables, constructed out of the gravitational and
matter phase-space variables themselves, instead of with respect to
the simple label $t$. The latter are called
{\it embedding variables} \cite{DIFIK1,DIFIK2,GCS}
and they can be considered as representing the location of a given
spacelike hypersurface in an embedding into a surrounding spacetime,
thus leading to the term {\it hypertime} for such formalisms. In a recent
paper \cite{CFS}, we looked at this method for the case of cylindrically
symmetric spacetimes, and succeeded in constructing hypertime
variables for those models. We will be extending that work in this paper
to the physically more interesting case of spherically symmetric black holes
with self-interacting matter fields.

We proceed as follows: In Section~\ref{hyper} we will
briefly describe the hypertime formalism, in the case when we have
boundary conditions. In particular we will indicate the conditions
needed for a hypertime formalism to be consistent. In Section~\ref{ssg}
we will describe the model being used, and construct the corresponding
ADM geometrodynamics. The important part of the calculation is
then found in Section~\ref{hbh}, in which we actually find a set
of hypertime variables for spherically symmetric gravity, and then
verify that they meet the conditions described in Section~\ref{hyper}.
Finally, in Section~\ref{disc}, we will briefly discuss a few points
that arise out of such a calculation, including quantization,
the connection to work by Unruh \cite{BILLNOTES} and to the work
on the semiclassical stability of the inner horizon of black holes, by
Hiscock \cite{MASSI} and others.

\section{Hypertime and Boundary Conditions}\label{hyper}

We will first need to extend the hypertime formalism
developed by Kucha\v{r} \cite{KK3}, which is originally
applicable only to geometrodynamics without boundary conditions,
to one which can handle such conditions. The modifications are,
in fact trivial, and we shall just sketch them. The method starts by finding a
canonical transformation in which we can divide the
standard local gravitational phase-space variables into
three classes \cite{KK3}
\begin{equation}{}\label{transform}
g_{ij},\pi^{ij} \rightarrow T^\mu, \Pi_\mu, Y_\mu \equiv \{ g^A, \pi_A \},
\end{equation}
where $\mu$ ranges from one to four and
$A \in \{1,2\}$.
$g^A$ and $\pi_A$ represent
the `true' dynamical variables of the theory discussed
above, which can easily be extended to include matter degrees of freedom
without changing our results. $T^\mu$ represent the corresponding
internal embedding coordinates  (describing the location of the
hypersurface in spacetime) with $\Pi_\mu$ being the corresponding momenta
(i.e.\ the energy-momentum densities). The superhamiltonian
and supermomenta will then generally become non-local functionals
of the new variables.
For geometrodynamics with boundary conditions,
we will also allow the hypertime variables to include an added discrete
number of boundary hypertime variables $T_{(\partial)}^\omega$, with
corresponding
momenta $\Pi_{(\partial)\sigma}$,  on top of  each spatial point hypertime
$T^\mu ({\bf x})$. These will usually be needed when non-vanishing boundary
terms
are converted to constraints via the method of parametrization of infinity, an
example
of which is in a recent paper \cite{KHOLE} by Kucha\v{r} on the
geometrodynamics
of Schwarzschild wormholes. We will therefore be interested in
a total Hamiltonian of the following form:
\begin{equation}{}\label{totham}
{\bf H} = \int d^3x \left( N ({\bf x}) {\cal H} ({\bf x})
             + N_i ({\bf x}) {\cal H}^i ({\bf x}) \right) +
 N_{(\partial)} {\cal H}_{(\partial)} + N_{i (\partial)}{\cal
H}^i_{(\partial)},
\end{equation}
where the Lagrange multipliers $N_{(\partial)}$ and
$N^i_{(\partial)}$ enforce extra ``boundary'' constraints,
\begin{equation}{}\label{bcons}
{\cal H}_{(\partial)} [ T^\mu, \Pi_\nu, T_{(\partial)}^\omega,
\Pi_{(\partial)\sigma},
Y_\rho ] = 0,\ {\cal H}_{(\partial) i}
[T^\mu, \Pi_\nu, T_{(\partial)}^\omega,  \Pi_{(\partial)\sigma},
Y_\rho ] = 0.
\end{equation}
We can let these constraints be arbitrary functionals of the hypertime and
dynamical phase-space variables (as they will usually be generated by
non-local canonical transformations acting on standard ADM boundary
terms). We further assume that the extra constraints have been chosen such that
they are preserved under evolution using the
total Hamiltonian, so the extended constraint algebra does not
generate any extra constraints.
To simplify the equations, and to make the similarity of the
calculation to that by Kucha\v{r} \cite{KK3} obvious, we will introduce a
special indexing
and summation convention. Upper case Greek characters will range over tensor
indices,
spatial location, and boundary terms, with summation being over the entire
range, so that
\begin{equation}{}\label{Esumex}
\Pi_\Lambda \dot{T}^\Lambda= \int d^3x \,\Pi_\mu ({\bf x})
  \dot{T}^\mu ({\bf x})
   + \Pi_{(\partial)\mu} \dot{T}_{(\partial)}^\mu,
\end{equation}
which generalizes the convention used by Kucha\v{r} to include the boundary
terms.
Functional derivative language will be used, with the assumption that we use
partial derivatives when operating on appropriate objects.
We will also denote time-space pairs of variables $[ N, N_i ]$ by $4$-vector
notation $N_\mu$.

With the above notation, we start off by writing down a solution, for
$\Pi_\Lambda$,
to the constraint system,
\begin{equation}{}\label{SolnPI}
\Pi_\Lambda = -P_\Lambda [ T^\Sigma, Y_\mu ],
\end{equation}
so that we have
\begin{equation}{}\label{solvcon}
{\cal H}^\Lambda [ T^\Sigma, -P_\Sigma [ T^\Delta, Y_\nu ],
Y_\mu ] = 0.
\end{equation}
Functional differentiation of this gives us the following
relations:
\begin{equation}{}\label{momproj}
\frac{\delta {\cal H}^\Lambda}{\delta T^\Sigma} =
{\cal H}^{\Lambda \Delta} \frac{\delta P_\Delta}{\delta T^\Sigma},
\ \frac{\delta {\cal H}^\Lambda}{\delta Y^\mu} =
{\cal H}^{\Lambda \Delta} \frac{\delta P_\Delta}{\delta Y^\mu},
\end{equation}
where we define
\begin{equation}{}\label{fundef}
{\cal H}^{\Sigma \Lambda} = \frac{\delta {\cal H}^\Sigma}{\delta \Pi_\Lambda}.
\end{equation}
The dynamics of the embedding variables are then given by
\begin{equation}{}\label{embdyn}
\dot{T}^\Lambda = N_\Sigma {\cal H}^{\Sigma \Lambda},
\end{equation}
The important point is to now assume that ${\cal H}^{\Sigma \Lambda}$ can
be inverted, so that we can use Equation~(\ref{embdyn}) to write the
lapse-shift
vectors $N_\Sigma$ in terms of the hypertime velocities $\dot{T}^\Lambda$.
This is a crucial condition, and signifies the fact that the $T^\Lambda$
variables do indeed measure the embedding of the spacelike hypersurface
into the surrounding spacetime.

It is then a simple matter to follow Kucha\v{r} \cite{KK3}, and
substitute the solution for $N_\Sigma$ into the Hamilton equations for the
dynamical variables
\begin{equation}{}\label{dyndyn}
\dot{g}^A ({\bf x}) =  N_{\Lambda} \frac{\delta {\cal H}^{\Lambda}}{\delta
\pi_A ({\bf x})},
\ \dot{\pi}_A ({\bf x}) = - N_{\Lambda} \frac{\delta {\cal H}^{\Lambda}}{\delta
g^A ({\bf x})},
\end{equation}
to get the following evolution equation:
\begin{equation}{}\label{evodyn}
\dot{Y}^\mu = [ Y^\mu, P_\Sigma ]_P \dot{T}^\Sigma,
\end{equation}
where $[,]_P$ denotes the Poisson bracket over just the $Y^\mu$ variables.
The central point of the hypertime approach is to assume that $Y^\mu$
can be considered to be a functional of the total hypertime $T^\Lambda$ that
specifies the position of a spacelike hypersurface, and
not of the label time $t$. The evolution of $Y^\mu$ between one
hypersurface `at'
hypertime $T^\Lambda_1$ and another one at hypertime $T^\Lambda_2$ should
therefore be independent of the hypertime
foliation path $T^\Lambda (t)$ between them.
This leads to the following hypertime functional Hamilton equation:
\begin{equation}{}\label{HHam}
\frac{\delta Y_\mu}{\delta T^\Lambda} = [ Y_\mu, P_\Lambda]_P.
\end{equation}
For the evolution to {\it actually} be independent of the path,
we need to sure that the functional derivatives that are defined
the above equation weakly ($\approx$) commute, i.e.
\begin{equation}{}\label{comder}
\frac{\delta}{\delta T^\Sigma} [ Y_\mu, P_\Lambda ]_P =
\frac{\delta}{\delta T^\Lambda} [ Y_\mu, P_\Sigma ]_P,
\end{equation}
when the constraints are satisfied.
It is then simple to show that this is equivalent to the following condition:
\begin{equation}{}\label{Conscon}
\frac{\delta P_\Lambda}{\delta T_\Sigma} -
\frac{\delta P_\Sigma}{\delta T_\Lambda}
+ [ P_\Lambda, P_\Sigma ]_P =0,
\end{equation}
which is the extension of the equivalent condition
discussed by Kucha\v{r} \cite{KK3} that one would naively
expect. It is a simple matter \cite{KK3} to show that this is directly
implied by the (weak) vanishing of the total constraint algebra
\begin{equation}{}\label{vanalg}
[ {\cal H}^\Lambda, {\cal H}^\Sigma ]_T \approx 0,
\end{equation}
by using the assumed invertability of ${\cal H}^{\Lambda \Sigma}$ (where the
bracket $[,]_T$ is over all variables). Indeed, all we need to do is use the
inverse of ${\cal H}^{\Lambda \Sigma}$, and the relations given in
Equation~(\ref{momproj}) to rewrite the commutator in terms of $P_\Lambda$,
giving Equation~(\ref{Conscon}). Thus it is easy to
absorb boundary constraints into the hypertime formalism,
and to arrive at the generalized
hypertime evolution equation given by Equation~(\ref{HHam}).

The structure of the boundary terms in Equation~(\ref{totham})
is very important. If,
instead of constraints at the boundary, we had non-vanishing
boundary contributions to
the total Hamiltonian, we would not have the simple equivalence
between our formalism
and that of Kucha\v{r}. In particular, the inversion of
Equation~(\ref{embdyn}) would be more
complicated, and the consistency of Equation~(\ref{HHam}) would no
longer be guaranteed by the constraint algebra
closure condition, Equation~(\ref{vanalg}).

So, we have reduced finding a consistent geometrodynamical formulation
down to finding $P_\Lambda$. This may, in general, be impossible for
full general relativity, but it still seems useful to investigate the hypertime
formalism for model systems with extra symmetry. Just such a case
was investigated in Ref.~\cite{CFS}, when no extra boundary terms were
needed to achieve the correct dynamics. Now we are ready to reduce
spherically symmetric gravity to a Hamiltonian of the form given by
Equation~(\ref{totham}), and construct the corresponding $P_\Lambda$
functionals.

\section{Spherically Symmetric Geometrodynamics}\label{ssg}

We will be interested in a charged spherically symmetric metric minimally
coupled to a set of interacting real scalar fields. The charge will allow for
Reissner-Nordstr\"om wormhole
solutions as well as the more usual black hole ones. The scalar fields will
emulate the important technical aspects of a general matter content, without
adding the complexities of extra gauge constraints. The correct
geometrodynamics
of such models were first formulated by Unruh \cite{BILLNOTES}, based on
a paper by Berger, Chitre, Moncrief and Nutku \cite{BCMN}. The model has
been investigated in depth by Hajicek {\it et
al}~\cite{SSFI,SSFII,PETREC,SSFIV}.
Recently the vacuum
geometrodynamics for spherical black holes have been investigated by
D. Louis-Martinez, J. Gegenberg and G. Kunstatter \cite{LMGK1}, with a
corresponding hypertime formulation in the paper by Kucha\v{r} \cite{KHOLE}.

We will write the spherically symmetric metric,
with charge $q$, in a form that will lead to an
action that mirrors the one for cylindrically symmetric metrics used in
Ref.~\cite{CFS}
\begin{equation}{}\label{sphermet}
ds^2 = \varphi^{-\frac{1}{2}} e^{2\gamma} \left(
\left(-\alpha^2 +\beta^2 \right) dt^2 +
2\beta dt\,dt +dr^2 \right)
+ \varphi \left( d\theta^2 + {\sin}^2 (\theta) d\phi^2 \right),
\end{equation}
where the functions $\varphi, \gamma, \alpha$ and $\beta$ are functions only of
the time $t$ and radial coordinate $r$, and not of the coordinates for the
surfaces
of spherical symmetry $\theta$ and $\phi$. Each 2-sphere has area
$4\pi \phi (t,r)$, and $\gamma (t,r)$ specifies a conformal factor on
a 2-D spacetime with coordinate $(t,r)$ and metric
\begin{equation}{}\label{2dmet}
{}^{(2)}ds^2 =  e^{2\gamma} \left(
\left(-\alpha^2 +\beta^2 \right) dt^2 +
2\beta dt\,dt +dr^2 \right).
\end{equation}
The functions $\alpha (t,r)$ and $\beta (t,r)$ are the lapse and shift
functions,
respectively, connecting one constant $t$ spacelike hypersurface to
another. We now couple the metric to $N$ real scalar fields, $f_{(i)}$,
with an arbitrary self-interaction potential $V(f_{(i)})$, giving an action
in the 2-covariant form
\begin{equation}{}\label{2dact}
S = - \int dt\, dr\, \sqrt{g} \left(
\frac{1}{4} \varphi R +\varphi \sum_{i=0}^{N} \nabla^a f_{(i)} \nabla_a f_{(i)}
+ \varphi^{\frac{1}{2}} V  + \frac{1}{2} q^2 \varphi^{-\frac{3}{2}}
- \frac{1}{2} \varphi^{-\frac{1}{2}} \right),
\end{equation}
where $g$, $R$, and $\nabla_a$ represent the metric determinant, scalar
curvature and covariant derivative respectively on the 2-D spacetime with
metric
given by Equation~(\ref{2dmet}).  There will be important boundary
corrections to this action that we will calculate later.

We will assume
that the space is asymptotically flat in either one or two regions, and that
$(t, \pm r)$ are the standard Minkowski coordinates in the respective
asymptotic
limits. If there is only one asymptotic region $r \rightarrow \infty$, then we
will
have a regular centre at $r=0$, and the spacelike hypersurfaces will have
topology ${\bf R}^3$. If we have two asymptotic regions $r \rightarrow \pm
\infty$,
then we will have a wormhole topology, and the spacelike hypersurfaces will
be 3-cylinders (${\bf R} \times {\bf S}^2$). In either case,
$\varphi$ tends asymptotically to $r^2$ for large $|r|$, with
corrsponding conditions on $\gamma$. For the matter fields, we
will assume that $f_{(i)} = o (|r|^{-1/2})$, and that
$V = o( |r|^{-2})$ for $|r| \rightarrow \infty$.

We can now write down the standard ADM \cite{ADM} formulation
of the model, with $t$ acting as foliation label. The total action,
up to boundary terms, takes the following form:
\begin{equation}{}\label{sphertac}
S = \int dt\, dr \, \left(
\pi_\gamma \dot{\gamma} + \pi_\varphi \dot{\varphi}
+ \sum_{i=0}^{N} \pi_{(i)} \dot{f}_{(i)}
- \alpha {\cal H} - \beta {\cal H}' \right),
\end{equation}
where $\pi_\gamma$, $\pi_\varphi$ and $\pi_{(i)}$ are the momenta
conjugate to $\gamma$, $\varphi$ and $f_{(i)}$ respectively.
${\cal H}$ and ${\cal H}'$ are the superhamiltonian,
\begin{equation}{}\label{sham}
{\cal H} = -2 \pi_\gamma \pi_\varphi - \frac{1}{2} \gamma_{,r}
\varphi_{,r} + \frac{1}{2} \varphi_{,rr}
+ e^{2\gamma} \left(
\varphi^2 V + \frac{1}{2} q^2 {\varphi}^{-\frac{3}{2}}
- \frac{1}{2} \varphi^{\frac{1}{2}} \right) + {\cal H}_f,
\end{equation}
and supermomentum
\begin{equation}{}\label{smom}
{\cal H}' = \pi_\gamma \gamma_{,r} + \pi_\varphi \varphi_{,r}
- \pi_{\gamma,r} + {\cal H}_f',
\end{equation}
and we have written ${\cal H}_f$ and ${\cal H}_f'$ for the $V=0$ part of the
superhamiltonian and supermomentum contributions for the scalar fields.
The momenta take the simple form
\begin{equation}{}\label{gmom}
\pi_\gamma  = -\frac{1}{2\alpha} \left(
\dot{\varphi} - \beta \varphi_{,r} \right),
\end{equation}
\begin{equation}{}\label{pmom}
\pi_\varphi  = -\frac{1}{2\alpha} \left(
\dot{\gamma} - \beta_{,r} - \beta \gamma_{,r} \right),
\end{equation}
\begin{equation}{}\label{fmom}
\pi_{(i)}  = -\frac{2\varphi}{\alpha} \left(
\dot{f}_{(i)} - \beta f_{(i),r} \right).
\end{equation}
The resulting constraints and momenta are very closely
related to those found for the cylindrically symmetric
system discussed in Ref.~\cite{CFS}.

\section{Hypertime for Black Holes}\label{hbh}

Now we are ready to take the results of Sections~\ref{hyper}
and~\ref{ssg} and formulate a hypertime geometrodynamics
for black holes and wormholes. We are directly motivated
by Ref.~\cite{CFS}, and therefore use basically the same
hypertime variable (which is derived from those originally
discovered by Kucha\v{r} for vacuum cylindrically
symmetric spacetimes \cite{KK1}). We define
\begin{equation}{}\label{defT}
T^\pm (r) = -2 \int_{r_L}^r \pi_\gamma (r') dr' \pm \varphi,
\end{equation}
with $r_L$ being the corresponding leftmost $r$ for the
wormhole and regular centre topologies.
The corresponding momenta are given by
\begin{equation}{}\label{defTP}
\Pi_\pm = \frac{1}{4} \frac{\partial}{\partial r}
\ln \left( \pm e^{-\gamma} \frac{\partial T^\pm}{\partial r}
\right) \pm \frac{1}{2} \pi_\varphi.
\end{equation}
These variables basically measure the rate of growth of the
area function $\varphi$ along appropriately
parametrized outgoing (increasing $r$) and incoming
(decreasing $r$) null geodesics, integrated along the spacelike
hypersurface. Indeed,  Equation~(\ref{gmom})
can be used to show that an outer apparent horizon occurs
at $T^+_{,r} = 0$, and an inner one at $T^-_{,r}=0$,
where our sense of outer and inner is respect to the
rightmost asymptotically flat region. For
a smooth, non-singular foliation, the hypertime
variables themselves are continuous and finite across these
horizons, and their momenta are well-defined on either size
of each horizon.

It is straightforward to show that
\begin{equation}{}\label{cantran}
(f_{(i)}, \pi_{(i)}, \gamma, \varphi, \pi_\gamma, \pi_\varphi)
\rightarrow ( f_{(i)}, \pi_{(i)}, T^\pm, \Pi_\pm ),
\end{equation}
is a canonical transformation but, unlike the same transformation
in Ref.~\cite{CFS}, it is {\it not} enough, on its own, to
form a hypertime transformation. The crucial missing factor
is the importance of the ADM boundary term in the
geometrodynamics of black holes and wormholes. We
can see this immediately, if we consider the paper by
Unruh~\cite{BILLNOTES}. In that paper, a specific foliation
and gauge choice
was used, corresponding to $\pi_\gamma=0$ and
$\varphi=r^2$ respectively. However, that
choice corresponds to $\dot{T}^\pm (r) =0$, and
therefore we need an extra hypertime variable to track the
evolving position of each leaf of the foliation, as $t$
increases into the future. From our discussion in
Section~\ref{hyper} we can easily guess that the
extra variable will be a boundary hypertime.

To construct the appropriate boundary terms, we will first write
down the transformed versions of Equations~(\ref{sham})
and~(\ref{smom}).
\begin{equation}{}\label{hsham}
{\cal H} = \Pi^+ T^+_{,r} - \Pi^- T^-_{,r} + {\cal H}_f
- \frac{1}{8} \left( 2V +\frac{q^2}{\varphi^2} - \frac{1}{\varphi} \right)
T^+_{,r} T^-_{,r}
\exp \left( \int_r^\infty dr'\, \left(
4\Pi^+ + 4\Pi^- - \frac{1}{2\varphi} \frac{\partial \varphi}{\partial r}
\right) \right),
\end{equation}
\begin{equation}{}\label{hsmom}
{\cal H}' = \Pi^+ T^+_{,r} + \Pi^- T^-_{,r} + {\cal H}_f',
\end{equation}
with $\varphi$ now simply being defined by Equation~(\ref{defT}).
However, if we smear these with the lapse and shift, to get the
total Hamiltonian, and vary, we get the following remaining
variations at each boundary:
\begin{equation}{}\label{bvar}
\delta \int dr' \left(
\alpha {\cal H} +\beta {\cal H}' \right)
= - \frac{1}{2} \delta \rho,
\end{equation}
where $\rho=\sqrt{|\varphi|}$, and we have solved the variation
equations everywhere else. This produces no extra variation
at a regular centre $\rho=0$, but we will need a correction term at
each asymptotic infinity. We find that we can cancel these
variations by adding the following correction to the total
Hamiltonian at each infinite limit of the $r$ integration:
\begin{equation}{}\label{bham}
\alpha_\pm H_{B\pm} = \alpha_\pm
\lim_{r \rightarrow \pm \infty} \frac{1}{2} \rho
 \int_r^\infty dr'\, \left(
4\Pi^+ + 4\Pi^- - \frac{1}{\rho}\frac{\partial \rho}{\partial r}
\right),
\end{equation}
where $\alpha \rightarrow \alpha_\pm$ as $r \rightarrow \pm \infty$.
We can understand these terms if we look at the effective
ADM mass at radius $r$,. The latter, scalar invariant,
quantity can be defined
by using the fact that we can always find a
coordinate system $(\bar{t}(t,r),\bar{r}(t,r))$
for a spherically symmetric geometry, at a single point,
in which the metric becomes that of a
Reissner-Nordstr\"om spacetime, with local
effective ADM mass $m(r)$ and charge $q$,
\begin{equation}{}\label{RNmet}
ds^2 = - \left( 1- \frac{2m(r)}{\bar{r}}
+\frac{q^2}{\bar{r}^2} \right) d\bar{t}^2
+ \left( 1- \frac{2m(r)}{\bar{r}} +
\frac{q^2}{\bar{r}^2} \right)^{-1} d\bar{r}^2
+ \bar{r}^2 \left( d\theta^2 + {\sin}^2 (\theta) d\phi^2 \right).
\end{equation}
If we calculate $\gamma$ by inverting Equation~(\ref{defTP}),
then we end up with the following very important equation:
\begin{equation}{}\label{masseq}
\Pi^+ +  \Pi^- = \frac{1}{4} \frac{\partial}{\partial r}
\ln \left( \rho - 2m + \frac{q^2}{\rho} \right).
\end{equation}
We can now use this relationship to get
\begin{equation}{}\label{ADMeq}
H_{B\pm} = \lim_{r \rightarrow \pm \infty}  m(r),
\end{equation}
by simple substitution in Equation~(\ref{bham}).
Thus $H_{B\pm}$ represents the standard ADM mass
correction, well known for spherical symmetry \cite{BILLNOTES,SSFI},
simply rewritten in terms of the new variables. The complete total
Hamiltonian will reproduce the equations of motion, but subject to
the condition that $\alpha$ is fixed at $r=\pm \infty$. This is good
enough for normal geometrodynamics but is not good enough
for hypertime geometrodynamics, from Section~\ref{hyper},  and
specifically for the following reason: To arrive at consistent hypertime
dynamics, the boundary terms must be constrained to vanish, which
would force $H_{B\pm}=0$ (corresponding to allowing variation of
$\alpha$ at infinity), and thus force the vanishing of
the ADM mass at each infinity.
The latter obviously cannot be allowed to happen
if we wish to study generic, non-vacuum, solutions. Kucha\v{r}, however,
has already addressed this problem in Ref.~\cite{KHOLE}, and indicated
that we can use the so-called method of {\it parametrization of
infinities} to transform our boundary terms to valid boundary constraints.

In our case, the method of parametrization of infinities starts by defining
two new variables, suggestively named $\tau_\pm$, with corresponding
momenta $\pi_{B\pm}$, and then using the following, extended, total
Hamiltonian:
\begin{equation}{}\label{bhtham}
{\bf H} = \int dr' \left( \alpha {\cal H} + \beta {\cal H}' \right)
+ \alpha_+ \left( \pi_{B+} + H_{B+} \right)
+ \alpha_- \left( \pi_{B-} + H_{B-} \right),
\end{equation}
where the final term is discarded when we have a regular centre.
This then gives us the following equations
of motion for the $\tau_\pm$ variables:
\begin{equation}{}\label{eomt}
\dot{\tau}_\pm = \alpha_\pm = \lim_{r \rightarrow \pm \infty} \alpha (r),
\end{equation}
and thus, on solutions, $\tau_\pm$ are just the proper times at
each infinity. We can now leave $\alpha$ free to vary, even at
infinity. We have the basic formalism described in
Section~\ref{hyper}, except that we still need to identify
a boundary hypertime variable, or variables.  Note that the
total constraint system is consistent, due to the fact that
${\bf H}$ does not contain $\tau^\pm$, and so
$\pi_{B\pm}$ are constants of the motion, as are the
ADM masses.

Will we first find the general solution to the local constraints, i.e.
to
\begin{equation}{}\label{bhcons}
{\cal H} [ T^\pm,\Pi_\pm,f_{(i)},\pi_{(i)} ] = 0,\ {\cal H}'
[ T^\pm,\Pi_\pm,f_{(i)},\pi_{(i)} ] = 0.
\end{equation}
If we use the effective ADM mass $m(r)$, via
Equation~(\ref{masseq}), we can
use the constraints to eliminate one functional
freedom in the solutions, and simplify the problem down to
\begin{equation}{}\label{Pp}
P_+ = \frac{T^-_{,r}}{4 \left( T^+_{,r} - T^-_{,r} \right)}
\frac{\partial}{\partial r} \ln \left(
\rho -2m + \frac{q^2}{\rho} \right)
+ \frac{{\cal H}_f'}{\left( T^+_{,r} - T^-_{,r} \right)},
\end{equation}
\begin{equation}{}\label{Pm}
P_- = - \frac{T^+_{,r}}{4 \left( T^+_{,r} - T^-_{,r} \right)}
\frac{\partial}{\partial r} \ln \left(
\rho -2m + \frac{q^2}{\rho} \right)
- \frac{{\cal H}_f'}{\left( T^+_{,r} - T^-_{,r} \right)},
\end{equation}
Where $P_\pm$ are the objects defined in Section~\ref{hyper}.
We can now rewrite the remaining part of the
constraints in terms of $m(r)$, giving
\begin{equation}{}\label{masspde}
\frac{\partial}{\partial r} \left( \rho -2m
+\frac{q^2}{\rho} \right) + 2 h_f \left(
\rho -2m +\frac{q^2}{\rho} \right)
+ \rho_{,r} \left( 2V\rho^2
+\frac{q^2}{\rho^2} -1 \right) = 0,
\end{equation}
where we have
\begin{equation}{}\label{hfdef}
h_f = \frac{1}{T^+_{,r}} \left(
{\cal H}_f+{\cal H}_f' \right)
- \frac{1}{T^-_{,r}} \left(
{\cal H}_f-{\cal H}_f' \right).
\end{equation}
It is a simple matter to solve this equation to get
\begin{equation}{}\label{solnm}
m(r) = \exp \left(  -2 \int_{r_L}^r dr\, h_f \right) M_L
+ \int_{r_L}^r dr' \left( h_f \left(
1+\frac{q^2}{\rho^2} \right) + V \rho \rho_{,r'} \right) \rho
\exp \left( -2 \int_{r'}^r dr''\, h_f \right),
\end{equation}
where $M_L=m(r_L)$ is a free integration constant.

We can now demonstrate that one, and only
one, of the new variables
$\tau_\pm$ can always be considered to be redundant,
by considering Equation~(\ref{solnm}) for each
topology. For a regular centre, $r_L=0$, we must
have $M_L=0$, and we then have the constraint
$\pi_{B+}+H_{B+}=0$, which, after solving
Equation~(\ref{bhcons}), corresponds to
\begin{equation}{}\label{regcen}
\pi_{B+}+\int_{0}^\infty dr \left( h_f \left(
1+\frac{q^2}{\rho^2} \right) + V \rho \rho_{,r} \right) \rho
\exp \left( -2 \int_{r}^\infty dr'\, h_f \right) = 0.
\end{equation}
Thus $\pi_{B+}$ is completely determined, and therefore
$\tau_+$ is a redundant variable. For the wormhole case,
$r_L=-\infty$, we have two boundary terms. The leftmost
constraint, using Equation~(\ref{solnm}), is simply
$\pi_{B-}+M_L=0$. This suggests that we make a simple
canonical transformation, and define $m_-= - \pi_{B-}$, with
corresponding momentum $\pi_{m-} = \tau_-$. Obviously $m_-$
is just the ADM mass measured at the left infinity.
The leftmost constraint is then just $M_L = m_-$. Taking this into
account, the remaining, rightmost, constraint becomes
\begin{equation}{}\label{wormie}
\pi_{B+} + \exp \left(  -2 \int_{-\infty}^\infty dr\, h_f \right) m_-
+ \int_{-\infty}^\infty dr \left( h_f \left(
1+\frac{q^2}{\rho^2} \right) + V \rho \rho_{,r} \right) \rho
\exp \left( -2 \int_{r}^\infty dr'\, h_f \right) = 0.
\end{equation}
We therefore find that $\pi_{B+}$ is determined once more,
and $\tau_+$ is again redundant.

The redundancy of $\tau_+$ suggests that we define
\begin{equation}{}\label{full}
T_{(\partial)} = \tau_+,\ \Pi_{(\partial)} = \pi_{B+},
\end{equation}
with $P_{(\partial)}$ being the corresponding
solution for $\pi_{B+}$, either from
Equation~(\ref{regcen}) or Equation~(\ref{wormie}).
In the wormhole case, the remaining variable
$m_-$ becomes one of the dynamical variables.
This gives us a set of $P_\Lambda$ functions, as
discussed in Section~\ref{hyper}, with a corresponding
division of our extended configuration space into dynamical
and hypertime variables. All that we need do now is
verify that we can invert the Hamilton equations for
the hypertime velocities $\dot{T}^\Lambda$ to get the
lapse and shift, and then the requirements specified
in Section~\ref{hyper} will be satisfied. It would a
horrendous algebraic problem to find the general
solution to Equation~(\ref{embdyn}), using the total
Hamiltonian that arises for the spherically symmetric
model, but we are lucky, as this is not something we
need to do. Continuity of the functional derivatives
in ${\cal H}^{\Sigma \Lambda}$, with respect to
the phase-space variables, ensures that there will
exist an inverse, in general, as long as we can find one
for a specific choice of those variables. We can simply
choose the foliation Unruh foliation and gauge choice
described above, and choose a vacuum configuration of
the fields. This results in the following equations:
\begin{equation}{}\label{vacuum}
0 = 2r \alpha \pm 2r \beta  + 2r\alpha_+ - 2 \int_0^r dr'\, \left(
\alpha +\alpha_+ \right).
\end{equation}
It is simple to verify that this gives us $\alpha=\alpha_+$ and
$\beta=0$, consistent with the boundary condition at $r=\infty$.
Thus the problem has an inverse in flat space and a
corresponding general solution. The conditions for consistency
of the hypertime formalism, given in Section~\ref{hyper} are
now satisfied, and we have achieved the aim of this paper.

\section{Discussion}\label{disc}
The Unruh formalism for spherically symmetric
spacetimes \cite{BILLNOTES} provides a valid Hamiltonian
geometrodynamics for collapsing matter. That formalism,
however, is for a specific choice of time (foliation) and
gauge and, up to now, it has been hard to tell whether
the manifest breaking of covariance causes any problems.
If we can show that it corresponds to one of the `paths' in
our hypertime formalism, then we may feel better about
using it as a basis for further (e.g. quantum) work. We also
provide a check on the new formalism itself.

The Unruh foliation and gauge corresponds to
\begin{equation}{}\label{billfol}
\dot{T}^\pm=0,\ \rho=r,\ \dot{T}_{(\partial)}=1.
\end{equation}
Equation~(\ref{evodyn}) then tells us that we
have
\begin{equation}{}\label{billchk}
\dot{f} = [ f, \int_{0}^\infty dr \left( h_f \left(
1+\frac{q^2}{r^2} \right) + V r \right) r
\exp \left( -2 \int_{r}^\infty dr'\, h_f \right) ]_P,
\end{equation}
for a single scalar field $f$, with the same equation for the
corresponding momentum. Thus we have a total reduced
Hamiltonian given by
\begin{equation}{}\label{billsHAM}
H = \int_{0}^\infty dr \left( h_f \left(
1+\frac{q^2}{r^2} \right) + V r \right) r
\exp \left( -2 \int_{r}^\infty dr'\, h_f \right),
\end{equation}
which is identical to that given by Unruh \cite{BILLNOTES},
once the correct matter fields are specified. Thus that formalism
corresponds to a specific path amongst all those possible in
the hypertime formalism, all of which produce equivalent dynamics.

This becomes important if we briefly examine the quantization
of our formalism. Under the general canonical quantization
rules, Equation~(\ref{SolnPI}) gives us the hypertime
functional Schr\"odinger equation
\begin{equation}{}\label{QuantHyper}
i \frac{\delta}{\delta T^\Lambda} \psi [ T^\Lambda,
Y_\mu ] = \hat{P}_\Lambda \psi [ T^\Lambda,
Y_\mu ].
\end{equation}
For the resulting quantum state to be independent of the
foliation condition, we need to satisfy the quantum equivalent to
Equation~(\ref{Conscon}), namely
\begin{equation}{}\label{quancom}
{\bf [ } \hat{\Pi}_\Lambda + \hat{P}_\Lambda,
\hat{\Pi}_\Sigma+ \hat{P}_\Sigma {\bf ] } = 0,
\end{equation}
on whatever the Hilbert space for the theory would be, as discussed
by Kuchar \cite{KK3}. If this condition holds, then the Unruh formalism
itself provides the basis for a quantum theory of black holes that
is foliation and gauge invariant. However, the hypertime formalism
allows us to use any foliation that can be specified by our embedding
variables. This can be seen to be important in the case of the Unruh
foliation, by the following problem: In that foliation, the spacelike
hypersurfaces avoid the apparent horizon, as discovered
by Hajicek \cite{SSFII,SSFIV}. This makes that approach inappropriate
for the study of the internal dynamics of black hole, and
of Hawking radiation. The hypertime variables, however, allow us
to work on foliations that extend into the black hole, through
the apparent horizon.

One feature of this calculation is that we have discovered the
importance of the mass function $m(r)$ in the quantum theory
of spherically symmetric spacetimes. The mass function is central
to understanding the semiclassical stability of the inner horizon in
black holes, as has been discovered by Hiscock \cite{MASSI}, Israel and
Poisson \cite{MASSIP}.
It may be possible that this formalism will therefore provide some insight into
that problem, and maybe even initial insight into a full quantum
version of the problem. Furthemore, the divergent behaviour of $m(r)$
found by the above authors could be important to the actual quantum
state of the black hole or wormhole.

\section{Conclusion}\label{conc}

We have developed a hypertime geometrodynamics for spherically
symmetric black holes and wormholes, as well as collapsing
matter field systems. The formalism is constructed around the
effective ADM mass function, and the embedding variables are suited
to work on the apparent horizons of wormholes and black holes.
The approach seems apt as a starting point for the quantization of
realistic black holes, via a hypertime functional Schr\"odinger equation.

\section*{Acknowledgments}

I would like to sincerely thank Bill Unruh for insightful comments on the
boundary constraint problem, Cynthia Foo for support, and
Malcolm MacCallum and Queen Mary and Westfield College, London,
for hospitality. The major part of this work was performed at the University
of British Columbia and was supported by the Natural Sciences and
Engineering Research Council of Canada.

%\bibliography{ref}

\end{document}